\documentclass[notitlepage]{revtex4-1}

\usepackage{graphicx}              
\usepackage[fleqn]{amsmath}   
\usepackage{mathtools}           
\usepackage{amsfonts}              
\usepackage{amsthm}

\begin{document}

\title{Recursion formula for the Green's function of a Hamiltonian for several types of
Dirac delta-function potentials in curved spaces}

\author{Fatih Erman}
\affiliation{Department of Mathematics, Izmir Institute of Technology, Urla
35430, Izmir, Turkey}

\email{fatih.erman@gmail.com}

%\author{{Fatih ERMAN}
%\\ \and
%{Department of Mathematics, Izmir Institute of Technology} \\ \and {Urla
%35430, Izmir, Turkey}
%\\
%{e-mail: fatih.erman@gmail.com}}

\date{\today}

\begin{abstract}
In this short article, we non-perturbatively derive a recursive formula for the Green's function associated with finitely many point Dirac delta potentials in one dimension. 
We also extend this formula to the case for the Dirac delta potentials supported by regular curves  embedded in two dimensional manifolds and for the Dirac delta potentials supported by two dimensional compact manifolds embedded in three dimensional manifolds.
Finally, 
this formulation allows us to find the recursive formula of the Green's function for the point Dirac delta potentials in two and three dimensional Riemannian manifolds, where the renormalization of coupling constant is required.  \end{abstract}


%\textbf{Keywords}: Dirac delta potentials, Green's functions, renormalization, Riemannian manifolds

\maketitle

\section{Introduction}

Dirac - delta potentials are considered as toy models in many different areas of physics. One of the best well-known example is the so-called Kronig-Penney model in condensed matter physics (see \cite{Demkov} and \cite{Belloni} for further applications in atomic and molecular physics, and  \cite{albeverio} for other applications). The subject attracts vast amount of attention in mathematics literature as well. One rigorous way of defining them is made by the theory of self-adjoint extension of symmetric operators \cite{albeverio}. Moreover, the study of Dirac delta potentials in more than one dimension provides a pedagogical framework of understanding the renormalization in a simpler context, namely in non-relativistic quantum mechanics \cite{Jackiw}.

In \cite{Besprosvany}, 
an explicit recursive formula for the Green's function $G^{(n+1)}(x,y)$ corresponding
to a Hamiltonian containing a sum of $n+1$ point Dirac delta-function potentials of arbitrary positions and strengths in
terms of $G^{(n)}(x,y)$ and the additional  $n+1$ point Dirac delta-function potential parameters has been found in one dimension. The main idea was based on the Lippmann-Schwinger equation written in the operator form: $G= G_0 - G_0 \, V \,  G $, where $G=(H-E)^{-1}$ and $G_0=(H_0-E)^{-1}$ are the full Green and free Green operators, respectively (they are also called resolvent). They are defined in \cite{Besprosvany} with a negative sign but it is just a matter of convention.  Here $V$ represents the interaction. In this case, $V$ is a sum of the Dirac delta function potentials with different coupling constants $\lambda_i$.

First of all,  Green's function $G^{(1)}(x',x)= \langle x' | G^{(1)} | x\rangle$ for one Dirac delta function potential is computed explicitly by solving the Lippmann-Schwinger equation iteratively and by summing the infinite Born series. Then,  the second delta function potential from the full Hamiltonian is separated and then the Green's function $G^{(2)}(x',x)$ is successively solved from the Lippmann-Schwinger equation. This pertubation expansion is similarly calculated and given explicitly by the equation (21) in \cite{Besprosvany}. Finally, one obtains the Green's function $G^{(n+1)}(x',x)$ in terms of $G^{(n)}(x',x)$ by induction. The formula is very useful from computational and numerical point of view since the number of calculations goes like $n$ whereas the number of calculations from the direct formula of Green's function goes like $n^2$.

However, this recursive formula is based on summing the Born series and its convergence is guaranteed if we impose that the coupling constants $U_i$ are sufficiently small. Hence the recursive formula found  in \cite{Besprosvany} is only valid under the above condition. 
However, we show here that we actually do not need to impose any condition for the convergence of the series. Instead we derive exactly the same result without consulting any perturbation expansion. Hence, one of the aim of this paper is to confirms the results obtained in \cite{Besprosvany} in a shorter, non-perturbative way so that we do not have to impose any condition on the convergence of the series. This is performed by writing the interactions as a sum of projection operators.

Moreover, after establishing the recursive formula, we also consider the Dirac delta potentials supported by curves embedded in two dimensional manifold and Dirac delta potentials supported by two dimensional compact manifolds embedded in three dimensional manifolds. These two models have been investigated in \cite{Burak1, Burak2}. The flat space version of these models are studied from the self-adjoint extension point of view and they are considered as a model for semiconductor quantum wires \cite{Exner1}. Here we show that the recursive formula for the Green's function of these systems can also be found similarly. Finally, we extend these recursion formulas to the point Dirac delta potentials in two and three dimensional Riemannian manifolds, where the renormalization is required.

\section{Recursive formula for Green's functions through a non-perturbative approach}
\label{Recursive formula for Green's functions through a Non-perturbative approach}

The Hamiltonian of the system for a particle interacting with $n+1$ Dirac delta interactions supported by finitely many points located at $a_i$ with strengths or coupling constants $\lambda_i$ in one dimension is formally given by 
\begin{equation}
H=-{\hbar^2 \over 2m} {d^2 \over d x^2} - \sum_{i=1}^{n+1} \lambda_i \delta(x-a_i) \psi(x) \;. 
\end{equation}
We can also write this Hamiltonian in the following abstract form in terms of the projection operator:
\begin{equation}
H = H_0 - \sum_{i=1}^{n+1} \lambda_i | a_i \rangle
\langle a_i | \;, \label{1d hamiltonian Dirac delta}
\end{equation}  
where $|a_i\rangle$ is the Dirac ket in the appropriate Hilbert space and $H_0$ is the free Hamiltonian (it could even include regular potentials). We
work out the resolvent formula of $H$ associated with $n+1$ Dirac delta potentials in terms of the resolvent formula associated with $n$ Dirac delta potentials. For that purpose, let us assume
that the two Dirac kets $| \psi \rangle$ and $| \chi \rangle$ are
related in such a way that the equality $(H-E) | \psi \rangle = |
\chi \rangle$ is satisfied. Then, we have
\begin{equation}
 \label{reseq} \left[ H_0 - E- \sum_{j=1}^{n+1} \lambda_j | a_j \rangle \langle
a_j | \right] | \psi \rangle = | \chi \rangle \;, \end{equation}
assuming complex number $E \not \in \mathrm{Spec}(H_{0})$. Separating the $n+1$ th term and get
\begin{equation}
 \left[ H_0 - E- \sum_{j=1}^{n} \lambda_j | a_j \rangle \langle
a_j | \right] | \psi \rangle = \lambda_{n+1} | a_{n+1} \rangle \langle
a_{n+1} | \psi \rangle + | \chi \rangle \;. 
\end{equation}
If we define $G^{(n)}=(H_0-\sum_{j=1}^{n} \lambda_j | a_j \rangle \langle
a_j |-E)^{-1}$ as the resolvent for the $n$ Dirac delta centers,  and act $G^{(n)}$ on both sides of the above equation from left, we find
\begin{equation}
|\psi \rangle = G^{(n)} | \chi \rangle + \lambda_{n+1} G^{(n)} |a_{n+1} \rangle \langle a_{n+1} | \psi \rangle \;. \label{psi}
\end{equation}
Projecting this onto
$\langle a_{n+1} |$, we obtain
\begin{equation}
 \langle a_{n+1} | \psi \rangle = {\langle a_{n+1} | G^{(n)} | \chi \rangle \over 1- \lambda_{n+1} \langle a_{n+1} | G^{(n)} | a_{n+1} \rangle} \;. 
 \end{equation}
Substituting this result into (\ref{psi}) we can read the resolvent formula for $n+1$ Dirac delta center problem, so we obtain the resolvent formula or equivalently resolvent kernel (Green's function) for $(n+1)$ centers in terms of the Green's function for $n$ centers and the parameters of $n+1$ th delta center:
\begin{equation}
G^{(n+1)}(x,y)= G^{(n)}(x,y) + {\lambda_{n+1} G^{(n)}(x,a_{n+1}) G^{(n)}(a_{n+1}, y) \over 1-\lambda_{n+1} G^{(n)}(a_{n+1},a_{n+1})} \;, \label{recursion formula in 1D}
\end{equation}
which is exactly same main result of the paper \cite{Besprosvany}. Here we have derived the same formula in a non-perturbative way and without assuming the validity of the convergence of the perturbation series.

We now consider the same problem with $n$ Dirac delta centers. If we act $G_{0}^{(n)}$ on (\ref{reseq}) for $n$ delta centers from left we find 
\begin{eqnarray}
|\psi \rangle = G^{(n)}_{0} \sum_{j=1}^{n} \lambda_{j} | a_{j} \rangle \langle
a_{j} | \psi \rangle + G^{(n)}_{0} | \chi \rangle \;.
\end{eqnarray}
By projecting $\langle a_i |$ from left, we find a matrix equation from which we can find $\langle a_{j}| \psi \rangle$. Substituting this into the above formula, we obtain the Green's function 
\begin{eqnarray}
G^{(n)}(x,y) = G_{0}^{(n)}(x,y) + \sum_{i,j=1}^{n} G_{0}^{(n)} (x,a_i) \Phi_{ij}^{-1}   G_{0}^{(n)} (a_j,y) \;,  
\end{eqnarray}
where
\begin{equation}
\Phi_{ij} =
\begin{cases}
\begin{split}  {1 \over \lambda_i} - 
 G^{(n)}_{0} (a_{i},a_{i}) 
 \end{split}
& \textrm{if $i = j$} \\
\begin{split}
-  \; G^{(n)}_{0}(a_{i},a_{j}) 
\end{split}
& \textrm{if $i \neq j$}.
\end{cases} \;.
\end{equation}
This formula reveals the fact that we have to make $n^2$ number of calculations to compute the Green's functions whereas the recursive formula only requires $n$ number of calculations.  
Hence, the recursive formula is computationally and numerically useful.

%%%%%%%%%%%%%%%%%%%%%%%%%%%%%%%%%%%%%%%%%%%%%%%%%%%%%%%%

\section{Recursion formula of Green's function for Dirac delta potentials supported by curves and surfaces}

Once the above non-perturbative calculations are established, we can also generalize the recursive formula for the Green's functions corresponding to the Dirac delta potentials supported by curves and surfaces. However, this generalization can only be seen easily from the above non-perturbative approach. It is rather difficult to see the result from the original perturbative calculations. In order to find the recursion formula we first need to recall some basic definitions of Dirac delta functions supported by curves and surfaces. 

Let $\Gamma:I \rightarrow \mathbb{R}^D$ be a regular curve in $\mathbb{R}^D$. Then, the Dirac delta function supported by a regular curve $\Gamma$ is defined as a distribution by 
\cite{Appel} 
\begin{eqnarray}
\langle \delta_{\Gamma}, \phi \rangle = \int_{\Gamma} \phi \; d s
\end{eqnarray}
for any test function $\phi$. Here, $d s$ is the integration element over the curve $\Gamma$ parametrized by $s$. Similarly, the Dirac delta function supported by a regular surface $S$ is defined by
\begin{eqnarray}
\langle \delta_{S}, \phi \rangle = \iint_{S} \phi \; d^2 S 
\end{eqnarray}
for any test function $\phi$. Here $d^2 S= \sqrt{g}  du d v$ is the integration element over the surface $S$ in local coordinates $(u,v)$ and $g$ is the determinant of the induced metric on $S$.

As is well-known, the point Dirac delta function located at $\mathbf{a}$ is defined through $\langle \delta, \phi \rangle = \phi(\mathbf{a})$, and from this we usually write the following formal expression 
\begin{eqnarray}
\int_{\mathbb{R}^D} d^D r \; \delta(\mathbf{r}-\mathbf{a}) \phi(\mathbf{r}) = \phi(\mathbf{a})
\end{eqnarray} 
so that we can use the symbol $\delta(\mathbf{r}-\mathbf{a})$ formally. Similarly, we can also deduce the formal expression for Dirac delta functions supported by a regular curve $\Gamma$ through the relation
\begin{eqnarray}
& & \langle \delta_{\Gamma}, \phi \rangle = \int_{\mathbb{R}^D} d^D r \; \delta_{\Gamma}(\mathbf{r}) \phi(\mathbf{r}) \cr & & \hspace{2cm}= \int_{\mathbb{R}^D} d^D r \; \left[\int_{\Gamma} d s \; \delta (x-x(s), y-y(s), z-z(s)) \right]\; \phi(\mathbf{r}) \;, \label{dirac curve}
\end{eqnarray}
where the parametric equations for the curve $\Gamma$ are given by $x=x(s)$, $y=y(s)$, and $z=z(s)$. The formal expression of the Dirac delta functions supported by a regular surface $S$ is similarly given by
\begin{eqnarray}
& & \langle \delta_{S}, \phi \rangle =\int_{\mathbb{R}^D} d^D r \; \delta_{S}(\mathbf{r}) \phi(\mathbf{r}) \cr & & \hspace{1.5cm} = \int_{\mathbb{R}^D} d^D r \; \left[\int_{S} d^2 S \; \delta (x-x(u,v), y-y(u,v), z-z(u,v)) \right]\; \phi(\mathbf{r}) \;. \label{dirac surface}
\end{eqnarray}
Furthermore, one can also define the Dirac delta function supported by a curve $\Gamma: I \rightarrow M$ embedded in a $D$ dimensional Riemannian manifold $(M,\tilde{g})$ and Dirac delta function supported by a isometrically embedded two dimensional compact Riemannian submanifold $(\Sigma,g)$ of a three dimensional ambient Riemannian manifold $(M,\tilde{g})$. Then, the above definitions (\ref{dirac curve}) and (\ref{dirac surface}) take the following forms
\begin{eqnarray}
\langle \delta_{\Gamma}, \phi \rangle = \int_{M} d_{\tilde{g}} \mu(x) \;  \delta_{\Gamma}(x) \phi(x) = \int_{M} d_{\tilde{g}} \mu(x) \; \left[\int_{\Gamma} d_{\tilde{g}} s \; \delta_{\tilde{g}} (x,\Gamma(s)) \right]\; \phi(x) \;, \label{manifold dirac curve}
\end{eqnarray}
\begin{eqnarray}
\langle \delta_{\Sigma}, \phi \rangle = \int_{M} d_{\tilde{g}} \mu(x) \;  \delta_{\Sigma}(x) \phi(x) = \int_{M} d_{\tilde{g}} \mu(x) \; \left[\int_{\Sigma} d_g \mu(x') \; \delta_{\tilde{g}} (x',\tilde{x}) \right]\; \phi(x) \;, \label{manifold dirac surface}
\end{eqnarray}
respectively. Here $d_{\tilde{g}} \mu(x')$ is the measure on the embedded submanifold at $x' \in \Sigma$ (it is found from the pull-back of the Riemannian volume element on $M$) and $d_{\tilde{g}} \mu(x)$ is the measure on the  Riemannian manifold at the point $x \in M$.

We will now consider two more models including Dirac delta potentials supported by curves and compact manifolds embedded in a higher dimensional Riemannian manifolds. They have been first  studied in \cite{Burak1, Burak2}. 
Let us first consider a generalized Schr\"{o}dinger operator with $n+1$ Dirac delta interactions, whose supports are arc-length parametrized non-intersecting closed curves $\Gamma_i$ of length $L_i$ embedded in a two dimensional Riemannian manifold $(M,\tilde{g})$, i.e.,
\begin{eqnarray}
\langle x | H | \psi \rangle= -{\hbar^2 \over 2m} \nabla_{\tilde{g}}^{2} \psi(x) - \sum_{i=1}^{n+1} {\lambda_i \over L_i} \int_{\Gamma_i} d_{\tilde{g}} s \; \delta_{\tilde{g}}(x,\Gamma_i(s)) \; \int_{\Gamma_i} d_{\tilde{g}} s \; \psi(\Gamma_i(s)) = E \psi(x) \;.
\end{eqnarray}
Similar to the problem for point Dirac delta potential in one dimension, we can write the above Hamiltonian in terms of projection operators:
\begin{equation}
H= H_0 - \sum_{i=1}^{n+1} {\lambda_i \over L_i} | \Gamma_i \rangle \langle \Gamma_i | \;,
\end{equation}
where $|\Gamma_i \rangle$ is the ket vector in the appropriate Hilbert space and defined through $\langle x | \Gamma_i \rangle=\int_{\Gamma_i} d_{\tilde{g}} s \; \delta_{\tilde{g}}(x,\Gamma_i(s)) $. However, this above Hamiltonian is exactly the same form as (\ref{1d hamiltonian Dirac delta}) except for the factor $\lambda_i$ is replaced by $\lambda_i/L_i$ and $|a_i \rangle$ is replaced by $\Gamma_i$. Then, the recursion relation for the Green's function becomes
\begin{eqnarray}
G^{(n+1)}(x,y)= G^{(n)}(x,y) + {(\lambda_{n+1}/L_{n+1}) G^{(n)}(x, \Gamma_{n+1}) G^{(n)}(\Gamma_{n+1}, y) \over 1-(\lambda_{n+1}/L_{n+1}) G^{(n)}(\Gamma_{n+1}, \Gamma_{n+1})} \;, \label{recursion formula curve in 2D}
\end{eqnarray}
where 
\begin{eqnarray} & & 
G^{(n)}(\Gamma_{n+1}, \Gamma_{n+1}) = \langle \Gamma_{n+1} | G^{(n)} |\Gamma_{n+1} \rangle \cr & & \hspace{2cm} =   \int_{M \times M} d_{\tilde{g}} \mu(x) \; d_{\tilde{g}} \mu(y) \;  \langle \Gamma_{n+1} |x \rangle \; G^{(n)} (x,y) \; \langle y | \Gamma_{n+1} \rangle \;.
\end{eqnarray}
Finally, we can also consider the Dirac delta interactions supported by a isometrically embedded two dimensional compact submanifold $\Sigma$ in a three dimensional ambient manifold $M$. The generalized Schr\"{o}dinger equation is given by 
\begin{eqnarray}
-{\hbar^2 \over 2m} \nabla_{\tilde{g}}^{2} \psi(\tilde{x}) -\sum_{i=1}^{n+1} {\lambda_i \over V(\Sigma_i)} \int_{\Sigma_i}  d_g \mu(x') \; \delta_{\tilde{g}} (x' , \tilde{x}) \int_{\Sigma} d_g \mu(x'') \psi(x'') =E \psi(\tilde{x}) \;,
\end{eqnarray}
where $V(\Sigma)$ is the volume of the submanifold $\Sigma$. This Hamiltonian can be formally written as a sum of projection operators
\begin{eqnarray}
H= H_0 - \sum_{i=1}^{n+1} {\lambda_i \over V(\Sigma_i)} |\Sigma_i \rangle \langle \Sigma_i| \;,
\end{eqnarray}
where $\langle \tilde{x} | \Sigma_i \rangle = \int_{\Sigma_i}  d_g \mu(x') \; \delta_{\tilde{g}} (x' , \tilde{x}) $. Since this is exactly the same form as in the previous two models, the result is immediate, i.e.,
\begin{eqnarray}
G^{(n+1)}(x,y)= G^{(n)}(x,y) + {(\lambda_{n+1}/V(\Sigma_{n+1})) G^{(n)}(x, \Sigma_{n+1}) G^{(n)}(\Sigma_{n+1}, y) \over 1-(\lambda_{n+1}/V(\Sigma_{n+1})) G^{(n)}(\Sigma_{n+1}, \Sigma_{n+1})} \;. \label{recursion formula manifold in 3D}
\end{eqnarray}

\section{Recursion formula of Green's function for Dirac delta potentials in two and three dimensional manifolds}

We can also extend the above recursion formula to the case when we have many point Dirac delta potentials in two and three dimensional Riemannian manifolds. However, this case requires the renormalization of the coupling constants \cite{point Dirac on manifolds1}. The renormalization is necessary for Dirac delta potentials only when codimension is two and three in contrast to the above cases, where codimension is one. In order to find the recursion formula for Green's functions, let us first shortly review the construction of the regularized Green's function and then the renormalization procedure given in \cite{point Dirac on manifolds1}. The Shr\"{o}dinger equation for a single particle moving in two and three dimensional Riemannian manifold $(M,g)$ and interacting with attractive point Dirac delta potentials $\delta_{g} (x,a_i)$ supported by a finite set of isolated points $a_i \in M$ is formally given by
\begin{equation} - {\hbar^2 \over 2m} \nabla_{g}^2 \psi(x) - \sum_{j=1}^{N} \lambda_j \delta_{g}(x, a_j) \psi(x) = E \psi(x) \;, \label{point delta Sch on manifolds}
\end{equation}
where $\nabla_{g}^2 = \frac{1}{\sqrt{\mathrm{det}(g)}}
\sum_{i,j=1}^{D} \frac{\partial}{\partial x^i}
\left(g^{i j} \, \sqrt{\mathrm{det} (g)} \;
\frac{\partial}{\partial x^j}\right)$ is the Laplace-Beltrami operator and $\delta_{g}(x,a_j)$ is defined formally from the relation $\int_{\mathcal{M}} d_{g} \mu(x) \;  \delta_{g}(x,a_i)  f(x)  = f(a_i)$, and $d_g \mu(x)$ is the Riemannian volume element. Moreover, we suppose that $a_i \neq a_j$ for $i \neq j$. Similar to the one-dimensional case, we can express the interaction term as a sum of the projection operators $|a_j \rangle \langle a_j|$, where the ket $|a_j \rangle$ is defined in the appropriate Hilbert space and $\langle x |a_j \rangle = \delta_g(x,a_j)$. The most natural regularization in Riemannian manifolds is the heat kernel function $K_t(x,y)$. Since $\lim_{t \rightarrow 0^+} K_t(x,y) = \delta_g(x,y)$ in the distributional sense, we will replace the point Dirac delta function by heat kernel so that we have regularized Hamiltonian. In the abstract formal form, we will solve the following inhomogenous Schr\"{o}dinger equation to find the regularized Green's function
\begin{equation}
\left[ H_0 - E- \sum_{j=1}^{n} \lambda_{j}(\epsilon) | a_{j}^{\epsilon} \rangle \langle
a_{j}^{\epsilon} | \right] | \psi \rangle = | \chi \rangle \;, 
\end{equation}
where $\langle x | a_{j}^{\epsilon} \rangle = K_{\epsilon/2} (x,a_j)$ and the coupling constants $\lambda_j$ are functions of the cut-off parameter $\epsilon$. Acting the operator $(H_0-E)^{-1}=G^{(n)}_{0}$ from the left of both sides, we obtain 
\begin{eqnarray}
|\psi \rangle = G^{(n)}_{0} \sum_{j=1}^{n} \lambda_{j}(\epsilon) | a_{j}^{\epsilon} \rangle \langle
a_{j}^{\epsilon} | \psi \rangle + G^{(n)}_{0} | \chi \rangle \;.
\end{eqnarray}
By projecting $\langle a_i |$ from left, we find a matrix equation from which we can find $\langle a_{j}^{\epsilon}| \psi \rangle$. Substituting this into the above formula, we obtain the regularized resolvent 
\begin{eqnarray}
G^{(n)\epsilon} = G_{0}^{(n)} + \sum_{i,j=1}^{n} G_{0}^{(n)} |a_{i}^{\epsilon} \rangle \Phi_{ij}^{-1}(\epsilon) \langle a_{j}^{\epsilon}| G_{0}^{(n)} \;,  
\end{eqnarray}
where
\begin{equation}
\Phi_{ij} (\epsilon) =
\begin{cases}
\begin{split}  {1 \over \lambda_i(\epsilon)} - 
\langle a_{i}^{\epsilon} | G^{(n)}_{0} | a_{i}^{\epsilon} \rangle
\end{split}
& \textrm{if $i = j$} \\
\begin{split}
-  \; \langle a_{i}^{\epsilon} | G^{(n)}_{0} | a_{j}^{\epsilon} \rangle
\end{split}
& \textrm{if $i \neq j$}.
\end{cases} \;.
\end{equation}
If we follow the same line of arguments as in the one dimensional Dirac delta potential, we have also 
\begin{eqnarray}
G^{(n+1)\epsilon}= G^{(n) \epsilon} + {\lambda_{n+1} (\epsilon) G^{(n)\epsilon} |a_{n+1}^{\epsilon} \rangle \langle a_{n+1}^{\epsilon} | G^{(n)\epsilon} \over 1-\lambda_{n+1}(\epsilon) G^{(n)\epsilon}(a_{n+1}^{\epsilon}, a_{n+1}^{\epsilon})} \;. \label{recursion cut-off point delta}
\end{eqnarray}
Here, we have $\langle x | G^{(n)}_{0} | y \rangle =G^{(n)}_{0}(x,y)= \int_{0}^{\infty} {d t \over \hbar} \; K_t(x,y) e^{t E/\hbar} $ and 
\begin{equation}
\langle a_{i}^{\epsilon} | G^{(n)}_{0} |a_{j}^{\epsilon} \rangle =G^{(n)}_{0}(a_{i}^{\epsilon},a_{j}^{\epsilon}) = \int_{M \times M} d_g \mu(x) \; d_g \mu(y) \; K_\epsilon(x,a_i) G^{(n)}_{0} (x,y) K_\epsilon(a_j,y) \;. 
\end{equation}
If we choose the coupling constants
\begin{eqnarray}
{1 \over \lambda_i(\epsilon)}= \int_{\epsilon}^{\infty}{dt \over \hbar} \; K_{\epsilon}(a_i,a_i) e^{-t \mu_{i}^2} \;,
\end{eqnarray}
for all $i=1, \ldots, n+1$, and and sandwich (\ref{recursion cut-off point delta}) between $\langle x |$ and $|y \rangle$, and then take the limit $\epsilon \rightarrow 0^+$, we get a recursion relation for the renormalized Green's functions
\begin{eqnarray}
G^{(n+1)} (x,y)= G^{(n)}(x,y) + {G^{(n)}(x, a_{n+1}) G^{(n)}(a_{n+1},y) \over \left( \Phi_{n+1 n+1} - \sum_{i, j =1}^{n} G_{0}^{(n)} (a_{n+1}, a_i) \Phi_{ij}^{-1} G_{0}^{(n)} (a_j, a_{n+1}) \right) } \; ,
\end{eqnarray}
where $-\mu_i^2$ is the bound state energy of the particle associated with the $i$th Dirac delta center in the absence of all other centers and $\Phi_{ij}$ is called renormalized principal matrix and given by
\begin{equation}
\Phi_{ij} =
\begin{cases}
\begin{split}  \int_{0}^{\infty} {d t \over \hbar} \; K_t(a_i,a_i) \left( e^{-t \mu_{i}^2/\hbar} -e^{t E/\hbar}\right)
\end{split}
& \textrm{if $i = j$} \\
\begin{split}
-  \int_{0}^{\infty} {d t \over \hbar} \; K_t(a_i,a_j) e^{t E/\hbar}
\end{split}
& \textrm{if $i \neq j$}.
\end{cases} \;.
\end{equation}
Here $\Re (E)<0$ in order to make the integral convergent and it can be analytically continued to the other regions of the complex plane. As a final remark, we must also emphasize that we can also find the recursion formula for many Dirac delta function potentials supported by curves in three dimensional manifolds, where the renormalization is also required. This could be done by following the same line of arguments given above and the result is formally the same.

\section{Conclusion}

In this paper, we have derived non-perturbatively an analytical recursive formula for the Green's function associated with the different kinds of Dirac delta function potentials in curved spaces. Hence, this work is an extension of the work \cite{Besprosvany} to the several dimensions and to the  different kinds of Dirac delta potentials. In contrast to the perturbative approach given in \cite{Besprosvany}, we have here derived the same form of the formula in a more elegant way and show that it has a generic form independent of the space where the Dirac delta potentials are embedded and their types.

\section*{Acknowledgments} 
I would like to thank O. T. Turgut, M. Gadella, and O. Pashaev for reading the manuscript and for useful discussions.

\end{document}